\documentclass[letter,reprint,amsmath,amssymb,aps,prd,floatfix]{revtex4-2}
\usepackage{graphicx}% Include figure files
\usepackage{dcolumn}% Align table columns on decimal point
\usepackage{bm}% bold math
\usepackage{hyperref}% add hypertext capabilities
\usepackage{slashed}
\usepackage{braket}
\usepackage{comment}
\usepackage[english]{babel}
\usepackage{graphicx} % Required for inserting images
\usepackage{xcolor}

\definecolor{bcolor}{rgb}{0.0,0.6,0.6}

\newcommand*{\hbb}{\mathbb{H}}
\newcommand*{\cbb}{\mathbb{C}}
\newcommand*{\bbb}{\mathbb{B}}
\newcommand*{\ibb}{\mathbb{I}}
\newcommand*{\id}{\text{id}}
\newcommand*{\acal}{\mathcal{A}}
\newcommand*{\bcal}{\mathcal{B}}
\newcommand*{\tcal}{\mathcal{T}}
\newcommand*{\Tr}{\mathrm{Tr}}

\newcommand*{\avg}[1]{\langle #1 \rangle}
\newcommand*{\sw}{\mathcal{S}}

\usepackage[normalem]{ulem}

\begin{document}

\title{Channel-State duality with centers}

%\date{\today}

\author{Simon Langenscheidt}
 \email{s.langenscheidt@physik.lmu.de}
\affiliation{
Arnold Sommerfeld Center for Theoretical Physics, Ludwig-Maximilians-Universit\"at München, Theresienstrasse 37, 80333 M\"unchen, Germany \\ Munich Center for Quantum Science and Technology (MCQST), Schellingstrasse 4, 80799 M\"unchen, Germany
}%

\author{Eugenia Colafranceschi}
 %\altaffiliation[Also at ]{}%Lines break automatically or can be forced with \\
%\author{Second Author}%
 \email{ecolafra@uwo.ca}
\affiliation{%
 Department of Physics and Astronomy, Western University, N6A 3K7, London ON, Canada\\
 Department of Physics, University of California, Santa Barbara, CA 93106, USA
}%

\author{Daniele Oriti}
\email{doriti@ucm.es}
\affiliation{
  Depto. de Física Teórica, Facultad de Ciencias Físicas, Universidad Complutense de Madrid, Plaza de las Ciencias 1, 28040 Madrid, Spain, EU \\
 Munich Center for Quantum Science and Technology (MCQST), Schellingstrasse 4, 80799 M\"unchen, Germany\\
 Department of Physics, Shanghai University, 99 Shangda Rd, 200444, Shanghai, P.R.China
}%

\begin{abstract}
    We study extensions of the mappings arising in usual channel-state duality to the case of Hilbert spaces with a direct sum structure. This setting arises in representations of algebras with centers, which are commonly associated with constraints, and it has many physical applications from quantum many-body theory to holography and quantum gravity. 
    We establish that there is a general relationship between non-separability of the state and the isometric properties of the induced channel. 
    We also provide a generalisation of our approach to algebras of trace-class operators on infinite dimensional Hilbert spaces.\\
\end{abstract}
\maketitle
%\tableofcontents  

\,\,\\
    %Alternative introductions

%%%%%%%%%%%%%%%%%%%%%    
    %    
%%%%%%%%%%%%%%%%%%%%%%
%\newpage

\section{Introduction}
\begin{figure}
    \centering
    \includegraphics[width=0.8\linewidth]{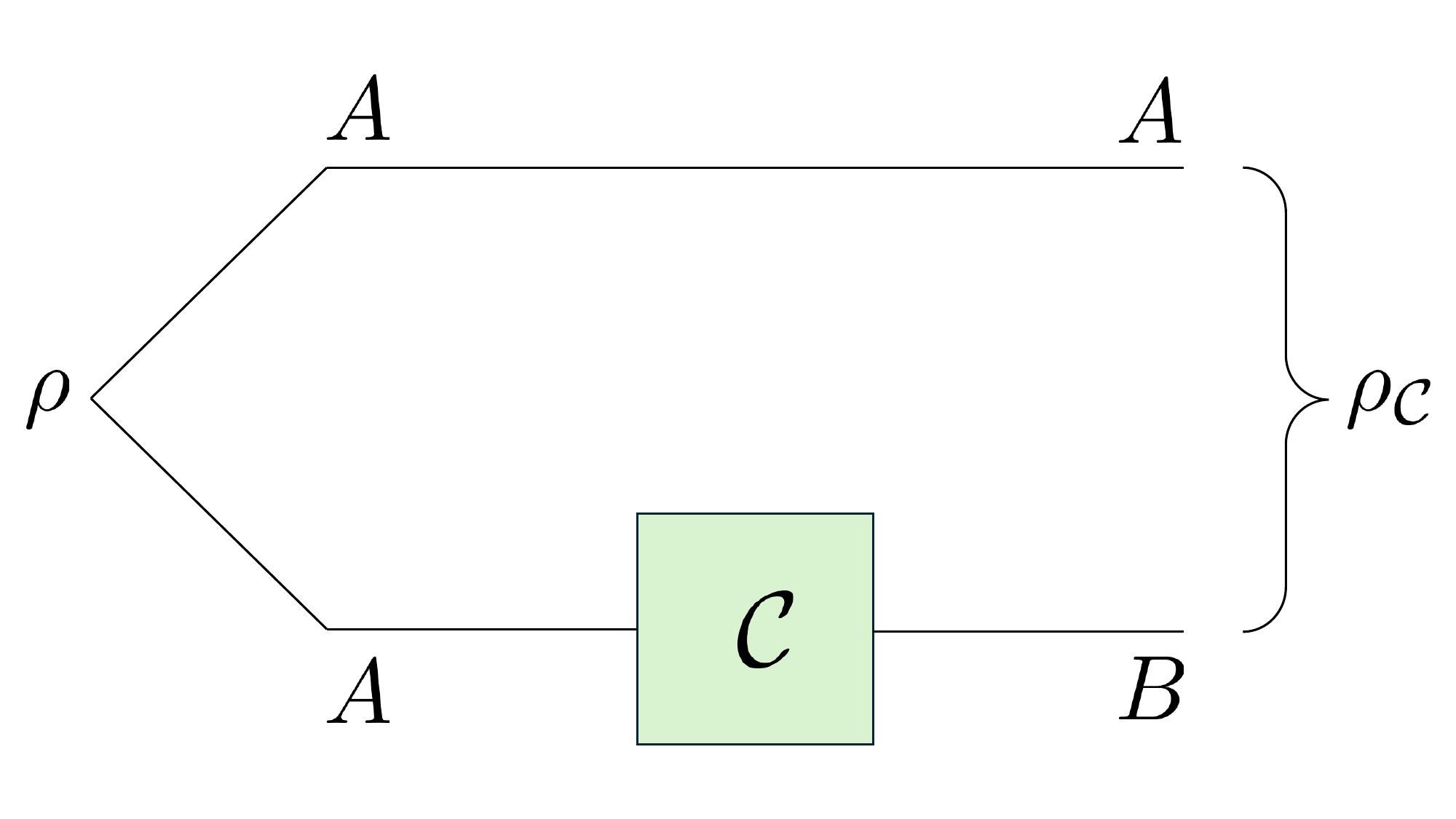}
    \caption{A quantum channel $\mathcal{C}$, which maps states from system $A$ to system $B$, is dual to a state $\rho_{\mathcal{C}}$ of the composite system $A+B$. This state is constructed by applying $\mathcal{C}$ to one half of a maximally entangled state $\rho$ shared between two copies of system A.}
    \label{fig:duality}
\end{figure}

In quantum information theory, the channel-state duality~\cite{jiang_channel-state_2013,majewski_comment_2013} refers to a correspondence between quantum operations (specifically, quantum channels) and quantum states of composite systems, which has profound implications at both mathematical and physical levels.  To be more specific, consider two systems, $A$ and $B$, described by Hilbert spaces $\hbb_A$ and $\hbb_B$, respectively, and a quantum channel $\mathcal{C}$ that maps states $\rho_A \in \bbb(\hbb_A)$ to states $\rho_B \in \bbb(\hbb_B)$. The channel-state duality establishes a correspondence between $\mathcal{C}$ and a specific state $\rho_{\mathcal{C}} \in \bbb(\hbb)$ over the extended Hilbert space $\hbb=\hbb_A \otimes \hbb_B$. The state $\rho_{\mathcal{C}}$ is constructed by applying $\mathcal{C}$ to a branch of a maximally entangled state $\rho \in \bbb(\hbb_A \otimes \hbb_A)$, as depicted in figure \ref{fig:duality}. 

In this setup, if $\rho_A$ represents an initial state and $\rho_B$ the final state, the quantum channel can be viewed as implementing a form of ``discrete dynamics''. In this perspective, the channel-state duality relates the kinematic properties of states to the dynamical properties of channels. Another relevant perspective, which is central to our work, arises from the fact that quantum states inherently define a quantum channel between two of their subsystems, as emphasized e.g. in \cite{ARRIGHI200426}.

In this work, we explore the channel-state duality in cases where the Hilbert space of the composite system $\hbb$ does not factorize over subsystems, but features a direct sum structure: $\hbb=\bigoplus_\mu \hbb^\mu_A \otimes \hbb^\mu_B$. This corresponds to situations where the algebras of observables for the systems possess a non-trivial center. In such cases, the channel-state duality can be used to characterize entanglement between subsets of quantum degrees of freedom, despite the fact that the Hilbert space lacks a clear tensor product decomposition, and so no unambiguous notion of subsystems is immediately available.
Such Hilbert space structures generically arise in systems which are subject to holonomic constraints, but are also of importance in other contexts.\\
In particular, a Channel-state duality of this more general kind has found use in tensor network models of holography\cite{colafranceschi_holographic_2022,hayden_holographic_2016,dong_holographic_2023,pastawski_holographic_2015}, where the systems $A$ and $B$ are seen as, respectively, the bulk and boundary of a spatially compact system, and state or operator reconstruction on boundaries are used as hallmarks of holographic behaviour. Indeed, this is a case in which the relevant Hilbert spaces are not, in general, composed of factors of the same size when they factorize over $A$ and $B$, or do not factorize at all~\cite{donnelly_decomposition_2012,freidel_corner_2023,bianchi_loop_2023}. Motivated by these models, we study  transport superoperators on bi- and tripartite systems subject to constraints which prevent the total Hilbert space to fully factorize, enforcing a structure of the form
\begin{equation}
    \hbb = \bigoplus_E \hbb_{I,E}\otimes\hbb_{O,E}.
\end{equation}
where labels $I$ and $O$ refer to the possible interpretation of the two subsystems as, respectively, "input" and "output", and where $E$ is the value of the constraint charge.
In particular, we ask the questions:   
\textit{Given such a setup, what is a sensible notion of complementary subsystems, and of a transport operator} $\tcal$ \textit{from a subsystem to a complementary one?} Further, \textit{Is there an associated version of channel-state duality for this direct sum setting?}

We will provide here a general framework to address these questions, leaving specific applications to future work.

The structure of this paper is as follows. In section \ref{Ch1}, we recapitulate the exact statement of Channel-state duality and present examples of non-factorised Hilbert spaces. In section \ref{Ch2}, we review the problem and introduce relevant concepts (e.g. algebraic subsystems, extension and partial trace maps). In section \ref{Ch3}, we study the case of transport operators on tripartite systems with no direct sum, and then extend the discussion to direct sums of tripartite systems. Section \ref{Ch4} presents the generalization of the Channel-State duality to the direct sum case. Section \ref{Ch5} discusses possible extensions of these notions to the infinite dimensional case.

\section{Channel-state duality}\label{Ch1}

Let us first give a simple example of the maps we consider, but in a much simpler context, i.e. %Our investigation can be motivated in the most elementary way by considering 
the simplest bipartite quantum system made from 2 qubits. 
For Bell states of 2 qubits, the action of an operator on one of the qubits can be expressed through an operator on the other, more specifically its transpose:\footnote{For the $\ket{\Psi^{\pm}} $ states, the corresponding operator is $X \sigma^1$, so multiplied by a Pauli matrix.}
\begin{equation}
    (X \otimes \ibb_B)\ket{\Phi^{\pm}} = ( \ibb_A \otimes \pm X^{t})\ket{\Phi^{\pm}}
\end{equation}
This is possible due to the maximal correlation between the two qubits in this state, which implies that acting on or measuring any property on one of the two is equivalent to performing some related action or measurement on the other. 
In the following, we study generalisations of this property, but with focus on sets of operators, and on how states with this property are selected. Already in this simple case, using this state, we can define a correspondence between operators on the two qubits that we call a \textit{transport superoperator}: a mapping 
\begin{equation}
    \tcal_{\Phi^\pm}:\bbb(\cbb^2_A) \rightarrow \bbb(\cbb^2_B)
\end{equation}
which takes any (bounded) operator on a subsystem and turns it into an operator on its complementary subsystem. Specifically,
\begin{equation}
    \tcal_{\Phi^{\pm}}(X) = \dim(\cbb^2_A)\cdot\Tr_A[ (X\otimes \ibb_B) \ket{\Phi^{\pm}}\bra{\Phi^{\pm}} ] \in \bbb(\cbb^2_B).
\end{equation}
One can verify easily in components that this does indeed give the transpose $\pm X^t$ of the operator $X$ for $\ket{\Phi^\pm}$. Written like this, it becomes clear that this is a particular case of channel-state duality~\cite{jiang_channel-state_2013,majewski_comment_2013}. Note that such a mapping is state-dependent. How well subsystems are mapped onto each other by its action depends both on how similar the subsystems are (in particular, the dimensions of the subsystem or their tensor product substructure) and how strongly correlated they are in the given state.

To be even more precise, the duality connects sets of $k$-positive operators on a bipartite system $\hbb = \hbb_A\otimes \hbb_B$, 
\begin{equation}
\begin{gathered}
    \mathcal{L}_k(\hbb) = \{ O\in \bbb(\hbb): \forall \ket{\psi}  \\
    \text{ of Schmidt rank } k \text{ or less}, \bra{\psi} O \ket{\psi} \geq 0 \}
\end{gathered}
\end{equation}
with the set of $k$-positive linear maps from $\bbb(\hbb_A) $ to $\bbb(\hbb_B)$, 
\begin{equation}
    \begin{gathered}
            BL_k(\bbb(\hbb_A),\bbb(\hbb_B)) = \{ \mathcal{T}:\bbb(H_A)\rightarrow \bbb(H_B): \\
            \ibb_{\cbb^k}\otimes \mathcal{T} \text{ positive} \}
    \end{gathered}
\end{equation}
in the form
\begin{equation}
    BL_k(\bbb(\hbb_A),\bbb(\hbb_B)) \cong \mathcal{L}_k(\hbb)  \qquad \forall k\leq \dim(\hbb_A).
\end{equation}
The most commonly used example of this duality is for $k=\dim(\hbb_A)$, where the set of operators becomes that of (unnormalised) states on $\hbb$, and the set of maps is that of completely positive(CP) ones. Then, restricting to normalised states/trace-preserving(TP) maps, one recovers a correspondence between channels and states. (For example, between $\ket{\Phi^\pm}$ and $\tcal_{\Phi^\pm})$.\\
Given a (1-)positive operator $\rho\in \bbb(\hbb)$, we can get the channel side of the duality in the form
\begin{equation}
    \tcal_\rho (X) = \Tr_A [(X\otimes\ibb_B)\Sigma(\rho)]
\end{equation}
where either $\Sigma(\rho) = \rho$ or $\Sigma(\rho) = \rho^{t_A}$ with the partial transpose $(-)^{t_A}$. The two options here may be referred to, respectively, as the Jamiolkowski-Pillis and the Choi mapping. The difference between them is that, while both establish an isomorphism for $k=1$, only the Choi map extends this to arbitrary $k$.  \\
The converse requires a noncanonical choice. Given a $k$-positive linear map $\tcal:\bbb(\hbb_A)\rightarrow\bbb(\hbb_B)$, we must choose a \textit{reference maximally entangled state} $\ket{\phi}\in\hbb$. With such a choice, we can define the associated k-positive operator $\rho_\tcal\in\bbb(\hbb)$
\begin{equation}
    \rho_\tcal := (\id_A\otimes\tcal)\ket{\phi}\bra{\phi}
\end{equation}
often referred to as the \textit{Choi matrix} of the channel. The reference state must be chosen maximally entangled in order for the resulting operator to carry the same information as the channel, but split between the subsystems $A,B$. (If it were for example a product state, then the resulting operator would only capture one point in the image of $\tcal$.) Different reference states $\ket{\phi}$ generally lead to different $\tcal\rightarrow  \rho_\tcal$ correspondences, although there are some exceptions (see e.g. \cite{jiang_channel-state_2013}). We will disregard these ambiguities in what follows, as we are primarily interested in the inverse of the correspondence, i.e. $\rho \rightarrow  \tcal_\rho$ . Still, we will point out, when needed, which choice we make to establish the $\tcal\rightarrow  \rho_\tcal$ direction of the generalised channel-state duality. This is particularly relevant because in the case of direct-sum Hilbert spaces, the correct analogue of a maximally entangled state is not immediately obvious. We will later see that for a sensible notion of channel-state duality for the direct-sum setting, it will simply consist of a maximally entangled state per sector.

We can see that the property of \lq \lq Bellness\rq\rq , i.e. maximal quantum correlation, of $\ket{\Phi^\pm}$ does not come into play in making $\tcal_{\Phi^\pm}$ a channel. Rather, regardless of the quantum correlations present in the state, as long as $\rho$ is normalised and completely positive, it induces a channel. In fact, we will demonstrate in the following that maximal correlation, translated in terms of maximal entanglement entropy, instead corresponds to an \textit{additional} property of the channel: it is \textit{isometric}, where isometry is defined in the Hilbert-Schmidt inner product on operators $ \langle X,Y\rangle_{HS} = \Tr[X^\dagger Y] $:
\begin{equation}
    \langle \tcal_{\Phi^\pm} (X),\tcal_{\Phi^\pm} (Y)\rangle_{HS,B} = \langle X,Y\rangle_{HS,A} ~.
\end{equation}
The notion of (information) transport superoperators associated with states, and their properties with respect to this inner product, can characterise the underlying state in a useful operational fashion.

With this in mind, we want to now present some examples of Hilbert spaces with a direct sum structure resulting from a nontrivial center, before discussing what impedes usual channel-state duality.

This setting is, as we stressed, quite generic. Consider two systems with a given series of eigenenergies $E$, each with eigenspace $\hbb_E$, and fix their total energy to $\mathcal{E} $. The corresponding Hilbert space is
    \begin{equation}
        \hbb_{\mathcal{E}} = \bigoplus_{E+E'=\mathcal{E}} \hbb_{E}\otimes\hbb_{E'}.
    \end{equation}
    So, we may generically see such direct sum spaces as indicating the imposition of a constraint in some larger (possibly factorized) Hilbert space.\\
    To give a couple more examples, we can first consider a spin system on a lattice $\Lambda$, described by a qubit on each site $v \in \Lambda$. If the dynamics is such that the total angular z-momentum $J = \sum_v \sigma^z_v$ is conserved, it makes sense to split the Hilbert space into its eigenspaces
    \begin{equation}
        \hbb_{\text{free}} = \bigotimes_{v\in \Lambda} \cbb_v^2 \cong \bigoplus_J \hbb_J
    \end{equation}
    In this case, the sectors are labeled by the total spin $J$ (whose range depends on the size of the lattice) and do not have a direct tensor product factorisation due to the global constraint, unlike the total (free) Hilbert space. If, however, the lattice is split into complements $\Lambda_{L|R}$, we can equally write them as
    \begin{equation}
        \hbb_J \cong  \bigoplus_{J_L+J_R = J} \hbb_{\Lambda_{L},J_L}\otimes \hbb_{\Lambda_{R},J_R}
    \end{equation}
    in direct analogy to the energy case before. In this case, then, the sector label $E$ is given by the total spin $J$ of the subregions.
    Contrast this with a case of \textit{local} constraints on each site, as in $\mathbb{Z}_2$ lattice gauge theory, with qubit on each \textit{link} $e\in \Lambda$. The system is subject to the Gauss constraints
    \begin{equation}
        G_v = \prod_{e\cap v}\sigma^x_e \stackrel{!}{=} \ibb_v 
    \end{equation}
    which are the lattice counterpart of the continuum Gauss constraint of gauge field theory and which impose a $\mathbb{Z}_2$ gauge invariance of states. The resulting gauge-invariant Hilbert space $\hbb_\Lambda$ does not have immediate spatial factorisation properties in terms of lattice sites, like in the example before. However, given a split of the lattice into complements $\Lambda_{L|R}$ along a set of links $S$, one can still give a \textit{direct-sum tensor split}
    \begin{equation}
        \hbb_{\Lambda} \cong \bigoplus_{E} \hbb_{\Lambda_{L,E}}\otimes\hbb_{\Lambda_{R,E}}
    \end{equation}
    in which the sectors are labeled by the set of $X_e$-eigenvalues on $S$,
    $E:=\{s_e\in \mathbb{Z}_2| e\in S\}$. Each of the tensor factors is simply the gauge invariant Hilbert space on the respective sublattice, subject to the boundary condition $X_e = s_e \,\forall\, e\in S$. Let us stress that  the same absence of local factorization is true in the continuum case and holds for any (field) system with gauge invariance\cite{donnellyEntanglementEntropyElectromagnetic2015,lin_comments_2018,freidel_corner_2023,freidel_quantum_2015}.\\
    These examples show also that as long as an a-priori tensor factorisation is present before imposing the constraints, one can keep said structure on the constrained subspace sector-by-sector in a direct-sum decomposition over the conserved charges.
    \\
    The lack of tensor factorisation makes unusable several technical and conceptual tools. The operator algebra associated with the full system no longer factorises nicely, and a clear notion of subsystems defined in terms of Hilbert subspaces also disappears. Further, the standard notion of entanglement, based on the deviation from a product state $\rho_A\otimes \rho_B $ on $\hbb_A\otimes \hbb_B$, is also no longer valid\cite{watrous_theory_2018,ma_entanglement_2016,donnelly_decomposition_2012,lin_comments_2018}, as no notion of a product state is available to compare to.\\
    We may of course choose to embed the direct sum in some bipartite system, with a factorized Hilbert space, but a choice of such embedding is highly nonunique and possibly unphysical. In fact, such a choice can be understood as equivalent to a choice of \textit{algebraic} subsystems for $A,B$, in the sense of choosing subalgebras $\acal_{A|B} \subset \acal=\bbb(\hbb)$\cite{zanardi_virtual_2001,hollands_entanglement_2018}. We deal with all of these issues in the following sections.

\section{The setting}\label{Ch2}
    Let us start from the algebra $\mathcal{A} = \bbb(\hbb)$ of operators\footnote{We will for simplicity work in this section with bounded operators on finite dimensional Hilbert spaces.} acting on a Hilbert space $\hbb$ of our choice. Then, select as subsystems two subalgebras $\mathcal{A}_I$,$\mathcal{A}_O$, to be later seen as inputs and outputs of the transport superoperator. These are understood as operations or observables of the subsystems in consideration. These two algebras are not necessarily a partition of $\mathcal{A}$, where a (bi)partition would be rather specified as the output being the algebraic complement (the commutant) of the input,
    \begin{equation}
        (\mathcal{A}_I)' = \mathcal{A}_O \qquad (\mathcal{A}_O)' = \mathcal{A}_I,
    \end{equation}
    so that the operators commuting with inputs are precisely the output operators. This structure, when present, captures the general properties we may expect of any operational definition of splitting of the system into two parts. \footnote{ Notice that, in particular, this is a pendent of properties in local QFT, as formalised by Haag duality, where subsystems are identified and distinguished by their localization on the spacetime manifold.}
    We have here labeled the two parts of the bipartition 'input' and 'output' in analogy to a quantum channel, but the labels could refer to any form of separation of subsystems (it could equally well be 'Left/Right', 'Inside/Outside', 'System/Measurement apparatus' or 'Alice/Bob', etc). For more about this algebraic perspective on subsystems, we recommend the review\cite{zanardi_virtual_2001}.\\
    In general, we have that $\mathcal{A}_I \cup \mathcal{A}_O \neq \mathcal{A}$, and, more importantly, we may have a nontrivial \textit{center} $\mathcal{Z} = \mathcal{A}_I\cap \mathcal{A}_O$, which consists of operators which commute with all others. If the center is trivial, i.e. consisting only of multiples of the identity $\lambda\ibb$, we have that $\mathcal{A}_I \cup \mathcal{A}_O$ factorises into $\acal_I\otimes\acal_O$. Its representations, then, also factorise into tensor products of Hilbert spaces. This is the simplest setting. In this special case, we can define subsystems as Hilbert subspaces in the tensor factorisation and extend subsystem operators uniquely: $X_I \mapsto X_I \otimes \ibb_O$\footnote{We assume here that the algebras contain a unit, so an identity operator.}; moreover, the  entanglement for pure states is well-defined and can be quantified, e.g. through von Neumann entropy. These properties do not generalise to the case with nontrivial center.\cite{ma_entanglement_2016,lin_comments_2018,casini_remarks_2014} \\
    In the case of nontrivial center, we have a \textit{commutative subalgebra} of $\mathcal{A}$. In fact, because the input and output systems commute with this center, we may seperate representations of the algebra into sectors labeled by eigenvalues of operators in $\mathcal{Z}$. This gives rise to the characteristic structure of the Hilbert space
    \begin{equation}
        \hbb = \bigoplus_{E} \hbb_E =  \bigoplus_{E} \hbb_{I,E}\otimes\hbb_{O,E} 
    \end{equation}
    in which the central operators have been diagonalised with eigenvalues given by $E$. A very concrete way to present this sort of algebra is through block matrices, where each block corresponds to a sector $E$ in the above decomposition. Lacking %These systems are characterised by not having 
    a clear notion of separable states, the simple notion of entanglement as non-product states does not hold up, but,  of course, this is not the end of the story, and for example entropies may still be calculated\cite{bianchi_typical_2019,ma_entanglement_2016,donnellyEntanglementEntropyElectromagnetic2015}. \\
    Still, the particular structure of the algebras and Hilbert spaces considered above indicates a natural way forward. 
    
    We still have a notion of subsystem in each sector due to factorisation, given by operators on input and output Hilbert spaces, in each of them. That is, there are associated subalgebras
    \begin{align}
       \bigoplus_{E} \bbb(\hbb_{I,E})\otimes\mathbb{I}_{O,E}  && && \bigoplus_{E} \mathbb{I}_{I,E}\otimes \bbb(\hbb_{O,E})
    \end{align}
    of operators on the individual subsystems for each sector. These are a consistent definition of complementary subsystems in the case of the Hilbert space structure above - they form subalgebras, have the correct commutant relation and the right center given by sums of identity operators in each sector, which represent the diagonalised operators from $\mathcal{Z}$. \\
    They also have unique extensions from the image of naive partial traces to the full algebra, and are the largest set to have this property. This means that, if we take the naive partial trace of an operator $X = \sum_{E,F} X_{E,F} \in \bbb(\hbb)$ (splitting into blocks over the different sectors), which is given by
    \begin{equation}
        \Tr_O [X] = \sum_{E} \Tr_{\hbb_{O,E}}[X] = \sum_{E} \Tr_{\hbb_{O,E}}[X_{E,E}]
    \end{equation}
    then we only keep the diagonal blocks in $X$, because there is no notion of a trace on the non-diagonal blocks. The reason for this is that, while there is a natural 'evaluation' or trace map on $V^\ast\otimes V$ for any vector space $V$, this is not true for 
    \begin{equation}
        \bbb(\hbb_E,\hbb_F) \cong \hbb_E^\ast \otimes \hbb_F
    \end{equation}
    when $E\neq F$ (or any vector spaces which are not equal). 
    Without providing such maps by hand (which amounts to a different choice of partial trace map), we can only arrive at operators of the above form by reducing to subsystems. 
    Similarly, if we want to extend some abstract subsystem operator, for example given by $U = \sum_{E,F} U_{E,F}  \in \bbb(\bigoplus_E \hbb_{I,E}) $ on the input subsystem, to the full system, then we would naively do so by extending it with 'identity operators' $\ibb_{O;E,F}$
    \begin{equation}
        i_I(U) := \sum_{E,F} U_{E,F} \otimes \ibb_{O;E,F},
    \end{equation}
    which however are only unambiguously defined, again, for $E=F$. This is simply the statement that the off-diagonal blocks will not have a clear notion of a diagonal, and certainly not of an 'identity', without prescribing it by hand. 
    In this sense, the choice of subsystem we indicate here is the only unambiguous one - for others, we would need to prescribe by hand extra data for defining any extension and restriction. 
To summarize, the natural way forward is to use the unique unambiguous definition of subsystems in each sector and extend it to the sum over sectors, to obtain information channels and a generalised channel/state duality for the whole system. What we do in the following is to show that this way forward can indeed be pursued, that it leads to a well-defined result, and that the resulting construction is, in the sense we clarified, the only natural one.

\section{Mappings on algebras with nontrivial centers}\label{Ch3}

        In order to define generalisations of the transport superoperator from before, we can follow the path indicated above, using a few ingredients:
    \begin{enumerate}
        \item Choices of input and output systems $\bcal_{I|O}$, e.g. $\bcal_{I|O} = \bigoplus_E \bbb(\hbb_{I|O,E}) $,
        \item Identifications/Injections $i_{I|O}:\bcal_{I|O} \hookrightarrow \acal $, whose images we identify as the complementary subsystems $\acal_{I|O}$,
        \item Conjugate partial trace maps $P\Tr_{I|O}:\acal \rightarrow \bcal_{I|O}  $ that reduce an operator on the full system to a subsystem,
        \item A mapping $\Sigma: \acal\rightarrow \acal$, usually  related to a density matrix $\rho$, e.g. $\Sigma(X) =  X \rho^{t_I} $.
    \end{enumerate}
    %If all of these are at hand, we can proceed with the same structure of mapping. 
    We will now go into detail about this construction in the case of a \textit{trivial center} at first, which corresponds to a system with simple tensor product factorisation in its Hilbert space. This will illustrate that the notion of transport operators is useful also in this simple case, and already shows the main behaviour of their properties, namely that there exists a 2-out-of-3 implication for purity of the state $\rho$, trace preservation and isometry of the mapping associated to it\footnote{This 2-out-of-3 property appears to be due to the relatively rigid way entanglement shows itself in pure states, as manifested through there being a (mostly unique) measure of entanglement for pure states, which is not the case for mixed states.}.
    After that, we show the generalisation to the case with multiple blocks/sectors or nontrivial center, and find that the same thing holds, but the conditions split per sector.

        \subsection{Trivial center}
            For starters, consider the C*-algebra $\acal = \bbb(\hbb)$ of bounded linear operators on a finite dimensional Hilbert space $\hbb$, equipped with the Hilbert-Schmidt inner product., together with a tripartition of $\hbb \cong \hbb_I\otimes\hbb_O\otimes\hbb_B$ into input, output and background spaces. In concrete cases, this induces subsystems $\acal_{I|O|B} $ and associated extension 
            \begin{equation}
                i_{I|O|B}: \bcal_{I|O|B} \hookrightarrow \acal
            \end{equation}
            and partial trace maps
            \begin{equation}
                P\Tr_{I|O|B}:\mathcal{A} \rightarrow \bcal_{I|O|B}
            \end{equation}
            which are, respectively, injective and surjective and, up to normalisation, inverses of each other. As before, we identify $\acal_{I|O|B} = Im(i_{I|O|B})$.\\
            We prescribe usually the subsystems as the obvious choice of subalgebras of $\bbb(\hbb)$:
            \begin{align}
                \bcal_I = B(\hbb_I) && \bcal_O = B(\hbb_O) && \bcal_B = B(\hbb_B).
            \end{align}
            In this concrete case, they are simply given by
            \begin{equation}
                i_I(X) = X \otimes \ibb_{BO} \qquad P\Tr_I(X) = \Tr_{BO}[X]
            \end{equation}
            and will from now on refer to $\bcal $ and $\acal$ interchangeably where there is no risk of confusion.\\
            In fact, the partial trace is best defined in terms of an adjoint to an extension.
             For example, for the bipartite case, it is the defining property of the partial trace that
            \begin{equation}
                \begin{gathered}
                    \avg{X_I\otimes\ibb_O,Y}_{HS} = \avg{X_I,\Tr_O[Y]}_{HS,I} \\
                    \forall \, X_I\in \bcal_I, \, Y\in \bcal_O
                \end{gathered}
            \end{equation}
            Using this relation, we can define more general partial trace and extension maps which share the same behaviour.
            
            These extension and partial trace maps may be used to create various kinds of transport maps from the input to the output system. In general, such a mapping will take the form
            \begin{gather}
                \tcal_\Sigma :\bcal_{I} \rightarrow \bcal_{O}\\
                    X \mapsto P\Tr_{O} [ \Sigma( i_I(X) ) ]
            \end{gather}
            where $\Sigma: \acal\rightarrow\acal$ is some linear mapping that twists the trivial extension-restriction operation. For $\Sigma=\id_{\acal}$, this gives a completely depolarising channel up to normalisation. We are interested in a twisting by multiplication with a density operator $\rho$ of the full system. These have the interpretation of first preparing the system in the state given by a density matrix $\rho$, acting on a subsystem $I$ with some operator and then looking at the results of that action in subsystem $O$. This gives an effective induced operator in $O$, and therefore provides a notion of 'operator transport' similar to the 2-qubit case that we discussed in the introduction. 
            
            The first concrete case we are interested in is the choice $\Sigma(X) = K\cdot X \rho$ with some positive constant $K$, which produces the Jamiolkowski-Pillis mapping
            \begin{equation}
                \tcal_\rho (X)  = K \, P\Tr_{O} [ i_I(X)\rho ].
            \end{equation}
            For the purpose of generality, we do not assume the density matrix has been trace-normalised and keep its appearance explicit in the following. This mapping is characterised by the relation in Hilbert-Schmidt inner products
            \begin{equation}\label{DefiningRelationJPCMapping}
                    \begin{aligned}
                        \avg{\tcal_\rho(X),Y}_O &= \avg{K i_I(X)\rho,i_O(Y)} \\
                        &= \avg{X,K\, P\Tr_I[\rho^\dagger i_O(Y)]}_I.
                    \end{aligned}
                \end{equation}
            The middle form here gives a clear interpretation of the inner products: we extend both $X$ from the input and $Y$ from the output to the full system, then take their inner product with the density matrix in between. Due to the cyclicity of the full trace on $\acal$, this is the same as the expression
            \begin{equation}
                \begin{aligned}
                    K\Tr_\hbb[i_O(Y)i_I(X)^\dagger
                \rho ] &= K \avg{ \rho, i_O(Y)i_I(X)^\dagger } \\
                &= K\avg{i_O(Y)i_I(X)^\dagger}_\rho
                \end{aligned}
            \end{equation}
            which is just, up to scaling, the expectation value of the operator given by $X^\dagger$ on the input subsystem and by $Y$ on the output subsystem, in the state $\rho$.\\

            We can, of course, change the twisting map $\Sigma$ to a different operation, but there is no uniquely compelling alternative. The perhaps most obvious alternative comes from a seemingly innocuous difference:
            the Choi mapping 
            \begin{equation}
                \tcal_\rho (X)  = K \, P\Tr_{O} [ i_I(X)\rho^{t_I} ].
            \end{equation}
            uses the \textit{partial transpose} of the state with respect to the subspace $\hbb_I$. 
            \iffalse
            This may be defined canonically using a series of isomorphisms and the swap operator:
            \begin{align}
                \text{vec}: &\mathbf{B}(\hbb) \rightarrow \hbb \otimes \hbb^\ast \\
                \mathcal{R}: & \hbb \otimes \hbb^\ast \rightarrow \hbb \otimes \hbb\\
                \sw_I: &\hbb_I \otimes \hbb_O \otimes \hbb_I \otimes \hbb_O \rightarrow \hbb_I \otimes \hbb_O \otimes \hbb_I \otimes \hbb_O \quad .
            \end{align}
            Letting the first be the canonical identification of linear operators with vectors, and the second the Riesz isomorphism, we can then define 
            \begin{equation}
                X^{t_I} := \text{vec}^{-1}( \mathcal{R}^{-1}(\sw_I\mathcal{R}(\text{vec}(X))) ) \quad .
            \end{equation}
            \fi 
            The Choi mapping has a number of more favorable properties compared to the Jamiolkowski-Pillis mapping. In particular, unlike the latter, the Choi mapping provides an isomorphism between the sets of CP maps $\hbb_I\rightarrow \hbb_O$ and of (unnormalised) states on $\hbb_I\otimes\hbb_O$. Also, for bipartite systems, the Choi mapping for a pure state $\ket{\phi}\bra{\phi}$ can always be written as
            \begin{equation}
                \tcal_\rho (X) = \Phi X \Phi^\dagger
            \end{equation}
            where the map $\Phi :\hbb_I\rightarrow \hbb_O$ has components $\bra{o}\Phi\ket{\iota} = \avg{\iota,o|\phi}$. This can be seen through 
            \begin{align}
                \bra{o}\tcal_\rho (X)\ket{\Tilde{o}} &= \sum_{i,\Tilde{i}} \bra{i}X\ket{\Tilde{i}} \bra{\Tilde{i},o} (\ket{\phi}\bra{\phi})^{t_I} \ket{i,\Tilde{o}}\\
                &= \sum_{i,\Tilde{i}} \bra{i}X\ket{\Tilde{i}} \bra{o}\Phi\ket{i} \bra{\Tilde{i}}\Phi^\dagger\ket{\Tilde{o}}.
            \end{align}
            \subsubsection*{Trace and isometry conditions}
            We will ask two important questions about this mapping: first, when it is a channel, and second, when it is an isometry (in the Hilbert-Schmidt sense).\\
            The former is a standard question, but the latter has an interesting new aspect to it: if the mapping is isometric, we can see the system itself as providing an 'information funnel' from input to output (this, in turn, has been used as a proxy of holographic behaviour in the literature).\\
            First, if $\rho\geq 0$, it is clear that the mapping is completely positive. Trace preservation amounts to 
            \begin{equation}
                \begin{gathered}
                    K\, P\Tr_{I}[\rho] = \ibb_{I} \\
                    \implies K = \frac{\dim(\hbb_I)}{\Tr[\rho]},\; P\Tr_{I}[\frac{\rho}{\Tr[\rho]}] = \frac{\ibb_{I}}{\dim(\hbb_I)} .
                \end{gathered}
            \end{equation}
            In other words, being a quantum channel fixes the overall normalisation of the mapping and also puts the requirement on the reduced input state that it must be flat.\\
            Isometry may be expressed easily as well, using swap operators\footnote{These simply take two factors in a tensor product and swap them, $\sw \ket{a}\otimes \ket{b} = \ket{b}\otimes \ket{a}$.} on the Hilbert spaces:
            \begin{align}
                &\avg{\tcal(X),\tcal(X) }_O \\
                &= \Tr_O[\tcal(X)^\dagger,\tcal(Y)] \\
                &= \Tr_{O^{\otimes 2}}[(\tcal(X)^\dagger\otimes\tcal(Y))\sw_O]\\
                &=K^2\,\Tr_{\hbb^{\otimes 2}}[(X^\dagger\otimes Y)(\rho\otimes\rho)\sw_O]\\
                &= K^2\,\Tr_{I^{\otimes 2}}[(X^\dagger\otimes Y) \Tr_{OB^{\otimes 2}}[(\rho\otimes\rho)\sw_O] ]
            \end{align}
            which translates into the requirement
            \begin{equation}
                K^2\, \Tr_{O^{\otimes 2}}[(\rho_{IO}\otimes\rho_{IO})\sw_{O}] = \sw_I.
            \end{equation}
            This in turn implies the two equalities (the second from multiplying the isometry condition by $\sw_I$)
            \begin{align}
                &K^2 \Tr[\rho]^2 e^{-S_2(\rho_O)} = D_I\\
                & K^2 \Tr[\rho]^2 e^{-S_2(\rho_{IO})} = D_I^2
            \end{align}
            expressed using the second Rényi entropy
            \begin{equation}
                e^{-S_2(\rho)} = \frac{\Tr[\rho^2]}{\Tr[\rho]^2} \quad .
            \end{equation}
            Combining these two leads to the general, normalisation-independent requirement
            \begin{equation}
                e^{-S_2(\rho_{IO})+S_2(\rho_{O})} = D_I \quad .
            \end{equation}
            The form of the exponent suggests looking for a subsystem inequality for Rényi entropies - however, it is known that such inequalities do not exist\cite{linden_structure_2013}\footnote{However, we might still use \textit{measured} Rényi entropies and mutual information\cite{scalet_computable_2021}. These have a known expression and satisfy nice properties as an analogue of the von Neumann mutual information.}. Still, in general these conditions fix $K^2\Tr[\rho]^2$ to be in the interval 
            \begin{equation}
                [D_I,D_I D_O]\cap [D^2_I,D^3_I D_O] = [D_I^2,D_I D_O].
            \end{equation}
            The minimum value $K = \frac{D_I}{\Tr[\rho]}$ is part of the trace condition above, while the maximum is $K = \frac{D_I}{\Tr[\rho]} \sqrt{\frac{D_O}{D_I}}$ is incompatible with being trace-preserving, in general. However, we should not preemptively choose the former value. Indeed, if we do, then the above conditions turn into
            \begin{align}
                & D^2_I e^{-S_2(\rho_O)} = D_I \implies S_2(\rho_O) = \log(D_I) \\
                & D^2_I e^{-S_2(\rho_{IO})} = D_I^2 \implies S_2(\rho_{IO}) = 0
            \end{align}
            which means that the reduced state $\rho_{IO}$ must be pure, and therefore the state must factorise $\rho=\rho_{IO}\otimes\rho_B$, and the reduced input (and output) state must be maximally mixed $S_2(\rho_O)=S_2(\rho_I) = \log(D_I)$. \\
            In other words, isometry (ISOM) and trace preservation (TP) imply purity of the state $\rho_{IO}$ (PURE).
            \begin{equation}
                \text{ISOM}\wedge \text{TP} \implies \text{PURE}
            \end{equation}
            
            However, this setup is too restrictive. On the other hand, the maximal value implies
            \begin{align}
                & e^{-S_2(\rho_O)} = \frac{1}{D_O} \implies S_2(\rho_O)= \log(D_O)\\
                & e^{-S_2(\rho_{IO})} = \frac{D_I}{D_O} \implies S_2(\rho_{IO}) = \log(D_O) - \log(D_I),
            \end{align}
            so once again the state reduced to the output system is maximally mixed. Now, however, the reduced state $\rho_{IO}$ no longer needs to be pure. So quite intriguingly, the mapping we propose cannot be a quantum channel and an isometry in general, unless the state used factorises in a nice way.\\
            Additionally, we may ask when the mapping we defined is unital. This gives an input-output swapped version of the trace preservation condition:
            \begin{equation}
                K\rho_O =\ibb_O \qquad \implies K = \frac{D_O}{\Tr[\rho]} \, , \frac{\rho_O}{\Tr[\rho]} = \frac{\ibb_O}{D_O}
            \end{equation}
            and we can again check when this is compatible with the mapping being isometric: we need $D^2_O \in [D_I^2,D_I D_O]$, but when $D_I \leq D_O$ this is only the case iff $D_I = D_O$. In that case, isometries are unitaries, and trace preservation and unitality are equivalent. Additionally, $K$ is fixed uniquely to the value $K = \frac{D_I}{\Tr[\rho]} = \frac{D_O}{\Tr[\rho]}$ and there is no other option than $\rho_{IO}$ being pure. \\
            We can frame this simple result as follows. If we fix a state $\rho$, then select manually input and output systems such that they are of equal size, then there is no way to have an isometry between the operator spaces from the Jamiolkowski-Pillis mapping if the state does not factorise into pure states.\\
            Even in this simple setting, operator transport has clear limitations in the multipartite case. The 'environment' or 'bath' $B$ generically makes it impossible for the mappings above to be isometric.\\
            
            To specialise this discussion, let us assume that the state $\rho_{IO} = \ket{\phi}\bra{\phi}$ is pure and we use either the Choi or Jamiolkowski-Pillis mapping (the requirements, for both of them, turn out to be the same). Then the isometry condition is
            \begin{equation}
                |K|^2 \Tr_{O^2}[\rho_{IO}^{\otimes 2}\sw_O] = |K|^2 \Tr_O[\rho_{IO}]^{\otimes 2} \sw_I = \sw_I \quad .
            \end{equation}
            This is simply the requirement of the reduced input state being flat:
            \begin{equation}
                \rho_I = \frac{\ibb_I}{|K|} = \frac{\ibb_I}{D_I}.
            \end{equation}
            This is precisely the condition we found before for trace preservation. So for pure states $\rho_{IO}$, trace preservation and isometry are in fact equivalent:
            \begin{equation}
            \begin{gathered}
                \text{PURE}\wedge\text{TP} \implies \text{ISOM} \\
                \text{PURE}\wedge\text{ISOM} \implies \text{TP} \quad .
            \end{gathered}
            \end{equation}
            This, together with the implication we found before, shows a 2-out-of-3 property of the Jamiolkowski-Pillis or Choi mappings. Phrased in terms of entanglement properties, we may say that for pure states, isometry holds precisely when the induced transport is trace preserving, or equivalently when the two subsystems are maximally entangled. \\
            
            We can also study the opposite case and ask what happens when the state $\rho_{IO}$ is seperable. In that scenario, we find (assuming a normalised $\rho$) that
            \begin{equation}
                \frac{\avg{\tcal_\rho(X),\tcal_\rho(Y)}_O}{\avg{X}_{\rho_I}\avg{Y}_{\rho_I}} = C_{seperable} = K^2 e^{-S_2(\rho_O) }  \quad ,
            \end{equation}
            which is, importantly, \textit{independent of $X$ and $Y$}. In simple terms, this is just the situation in which the inner product in $O$ factorises between $X$ and $Y$.  Of course, the same thing happens in the maximally mixed case - if $\rho_{IO} = \frac{\ibb}{D_I D_O}$ in the standard setup, then the above formula holds for maximal entropy of $\rho_O$.
            This suggests that, while the isometry condition indeed seems to favour entangled states, it also disfavours mixed states, generally.\\
            We note that already  in \cite{antipin_channel-state_2020}, it had been shown that even for mixed states one can link separability with properties of the induced transport superoperator. It would be interesting to extend these considerations to our more general setting, but we leave this for future work.

            \subsubsection*{An example}
            We illustrate, for concreteness, the Choi mapping on the classic Werner states on 2 qubits
            \begin{equation}
                \rho = p \Psi^- + (1-p) \frac{\ibb}{4} 
            \end{equation}
            where the Bell state $\Psi^- = \ket{\psi^-}\bra{\psi^-}$ is maximally entangled, and so the mapping is expected to give isometry. The Choi map (here for $K=2$) is linear in the state $\rho$, and, for the Bell state alone, induces a conjugation by the 2nd Pauli matrix:
            \begin{equation}
                \tcal_{\Psi^-}(X) = \sigma_2 X \sigma_2
            \end{equation}
            Therefore, the Werner states induce the superoperator
            \begin{equation}
                \tcal_\rho (X) = p \sigma_2 X \sigma_2 + (1-p)\frac{\ibb_O}{2} \quad .
            \end{equation}
            The isometry condition can therefore be checked directly:
            \begin{equation}
                \avg{ \tcal_\rho (X) , \tcal_\rho (Y) } = p^2 \avg{X,Y}_I + \frac{1-p^2}{2} \Tr_I[X]\Tr_I[Y] 
            \end{equation}
            and we can see that isometry only holds for the pure case $p=1$; it is not a consequence of entanglement by itself, but rather of entanglement together with purity. 
            
            \subsubsection*{Isometry degree of the average state}

            We can achieve a generic understanding of the tripartite case by employing Page-type averaging arguments\cite{page_average_1993,bianchi_typical_2019}. We can in principle just consider a random pure state $\rho= \ket{\psi}\bra{\psi}$ of the full system and compute quantities according to the unitary average $\avg{-}_U$, where states are given as $U\ket{\psi_{\text{ref}}}$. Then we can check the isometry condition in the average as well:
            \begin{equation}
                \begin{gathered}
                    \avg{|K|^2\, \Tr_{O^{\otimes 2}}[(\rho_{IO}\otimes\rho_{IO})\sw_{O}]}_U\\ 
                =
                |K|^2\, \Tr_{OB^{\otimes 2}}[\avg{\rho^{\otimes 2}}_U \sw_{O}] \quad , 
                \end{gathered}
            \end{equation}
            and use the result (found via Schur's theorem for the permutation group on the two copies of the system):
            \begin{equation}
                \begin{aligned}
                    \avg{\rho^{\otimes 2}}_U &= \int_{\mathcal{U}(D)} d\mu_{Haar}(U) \;(U^\dagger \rho U)^{\otimes 2} \\
                    &= \frac{\ibb_{\hbb\otimes\hbb}+\sw_{\hbb\otimes\hbb}}{D(D+1)} \quad .
                \end{aligned}
            \end{equation}
            This means that on average the left side of the isometry condition becomes
            \begin{equation}
                \frac{|K|^2 D_O^2 D_B }{D(D+1)} (\sw_I + \frac{D_B}{D_O} \ibb_{\hbb\otimes\hbb})
            \end{equation}
            which shows two conditions which must hold on average:
            \begin{equation}
                r= \frac{D_B}{D_O} << 1
                \qquad
                |K|^2 = \frac{D(D+1)}{ D_O^2 D_B } \approx r D_I^2 D_O \quad .
            \end{equation}
            So we can see already that only small environments allow for the average state to still give rise to isometries. This is unsurprising: in that scenario, a typical reduced state $\rho_{IO}$ is close to being pure. We can again take traces of this expression with $\ibb_I$ and $\sw_I$ to find 
            \begin{equation}
                D_I \stackrel{!}{=}1 \quad .
            \end{equation}
            We interpret this as follows. In order to have a system whose average pure state gives rise to an isometric map, the system sizes must follow the above conditions. Of course, if we restrict the average to a smaller class of states, we might find more lenient conditions. For example, we may only work with states of the form
            \begin{equation*}
                \rho = \Pi^\dagger \ket{\psi}\bra{\psi}\Pi
            \end{equation*}
            with some projector $\Pi:\hbb_B\rightarrow P\subset \hbb_B$ to a subspace of the environment, suitably extended to the full system. This essentially restricts the environment into a class of states. Then, the above calculation goes through as before, but replacing $D_B$ by $\Tilde{D}_B = \dim(P)$. Such a projection can then make the first condition superfluous by choosing $P$ to be small enough. So better knowledge of the state of the environment makes the effective Choi map more isometric. If we also want to check for trace preservation in this setting, we get the condition
            \begin{equation}
                \begin{gathered}
                    \frac{K D_O \Tilde{D}_B}{D} = 1 \longleftrightarrow K_{TP} = D_I \\ K_{Isom} = \sqrt{\Tilde{D}_B} D_I \quad .
                \end{gathered}
            \end{equation}
            So we have in fact that once again, choosing a small $P$ makes trace preservation and isometry nearly equivalent. So by either making the environment small, or choosing its coupling to the system to be small, or by assuming strong knowledge of the system (for example assuming it to be in a pure state, making $\Tilde{D}_B = 1$), we can find isometries in the tripartite case.

            \subsection{Nontrivial center}
                Let us first discuss the bipartite case.
                Consider an algebra $\acal$ with representation space $ \hbb $, pre-selected subsystems $\acal_{I|O}$, such that $(\acal_I)' = \acal_O$, but with nontrivial center\footnote{This assumes an extension map and associated partial trace operation have been chosen.} $\mathcal{Z}=\acal_I \cap \acal_O$. The case of interest to us is that of Hilbert spaces of the form
                \begin{equation}
                    \hbb = \bigoplus_E \hbb_{I,E}\otimes\hbb_{O,E} 
                \end{equation}
                with the full algebra $\acal = \bbb(\hbb)$, and subsystem algebras $\bcal_{I|O}  =\bigoplus_E \bbb(\hbb_{I|O,E})$. In this sector-split Hilbert space setting, extension and partial trace operations are defined sector-wise.
                \begin{align}
                    i_I(X) = \sum_E X_E \otimes \ibb_{O_E} \qquad P\Tr_I[X] = \sum_E \Tr_{O_E}[X_E] \quad , 
                \end{align}
                which are adjoints to each other under the Hilbert-Schmidt scalar products on the algebras. We also identify $\acal_{I|O} $, the true subsystems, as the images of $\mathcal{B}_{I|O}$ under the extension maps. In practical terms, any operator that may be reached by partial tracing needs to be in $\mathcal{B}_{I|O}$. Similarly, any operator that is obtained from extending one in $\mathcal{B}_{I|O}$ must be in $\acal_{I|O}$. \\

                We may once again define a Jamiolkowski-Pillis (or Choi with partial transpose) mapping via the property \ref{DefiningRelationJPCMapping}
                which is also fulfilled in the case of trivial center. We allow ourselves to rescale this mapping again by a constant K:
                \begin{equation}
                \begin{aligned}
                    \tcal_\rho (X)  &= K \, P\Tr_{O} [ i_I(X)\rho ] \\
                    &= \sum_E K c_E \, \Tr_{I_E} [ (X_{E,E} \otimes \ibb_{O_E})\rho_{E,E} ] \quad ,
                \end{aligned}
                \end{equation}
                where we decompose the state as
                \begin{equation}
                    \rho = \sum_{E,\Tilde{E}} \sqrt{c_E c_{\Tilde{E}}} \rho_{E,\Tilde{E}} \quad , 
                \end{equation}
                with $\Tr[\rho_{E,\Tilde{E}}] = \delta_{E,\Tilde{E}}$ and $c_E = \Tr_E[\rho] \geq 0$, $\sum_E c_E = 1$. \\
                
                For the tripartite case, we can proceed analogously. We assume: 1) a Hilbert space structure
                \begin{equation}
                    \hbb = \sum_{E} \hbb_{I,E}\otimes\hbb_{O,E}\otimes\hbb_{B,E}  \quad ;  
                \end{equation}
                2) input/output algebras
                \begin{equation}
                    \bcal_I = \bigoplus_E B(\hbb_{I,E}) \qquad \bcal_O = \bigoplus_E B(\hbb_{O,E}) \quad ;
                \end{equation}
                3) the mapping
                \begin{equation}
                    \begin{aligned}
                        \tcal_\rho (X)  &= K \, P\Tr_{O} [ i_I(X)\rho ] \\
                        &= \sum_E K c_E \, \Tr_{I_E B_E} [ (X_{E,E} \otimes \ibb_{O_EB_E})\rho_{E,E} ] \quad ,
                    \end{aligned}
                \end{equation}
                and investigate about trace preservation and isometry.\\
                Trace preservation is just the property
                \begin{align}
                     \Tr_{B_EO_E}[\rho_{E,E}] = \frac{\ibb_{I_E}}{D_{I_E}} && && c_E = \frac{D_{I_E}}{K} \quad .
                \end{align}
                \\
                Identifying isometry is made easier by the aforementioned relation \ref{DefiningRelationJPCMapping}, which entails that the adjoint to $\tcal_\rho$ is (unsurprisingly) given by
                \begin{equation}
                    \tcal^\ast_\rho (X) = P\Tr_I(\rho^\dagger i_O(X)) \quad .
                \end{equation}
                Letting $\sigma_E = \Tr_{B_E}[\rho]$ and rescaling our definitions by $K$, we obtain the isometry condition
                \begin{align}
                    &(\tcal^\ast\circ \tcal)(X) \\
                    &= \sum_E |K|^2\, \Tr_{O_E}[(\ibb_{I_E}\otimes \Tr_{I_E}[(X_E\otimes \ibb_{O_E})\sigma_E])\sigma_E^\dagger]\\
                    &= 
                    \sum_E |K|^2\sum_{a,b,c,d}
                    \avg{b|X_{E,E}|a} \cdot \\
                    &\cdot \ket{c}\bra{d} \cdot\Tr_{O_E}[
                    \avg{a|\sigma_E|b} \avg{c|\sigma_E|d}
                    ]\\
                    &\stackrel{!}{=} X = \sum_E
                    \sum_{a,b,c,d}
                    \avg{b|X_{E,E}|a} \cdot \ket{c}\bra{d} \cdot \delta_{a,d}\delta_{b,c} \quad ,
                \end{align}
                where we choose some orthonormal basis of $I_E$ labeled by $a,b,c,d$ in the last line. This leads directly to the condition
                \begin{equation}
                    |K|^2 \, \Tr_{O^2_E}[\sigma^{\otimes 2}_E \sw_{O_E}] = \sw_{I_E},
                \end{equation}
                as before. Notice however now that we use the same prefactor for \textit{all} sectors, meaning the requirement hinges more on the properties of the reduced states $\sigma_E$. Therefore the sector-wise condition
                \begin{equation}
                    e^{-S_2(\sigma_E)+S_2( (\sigma_E)_O )} = D_{I_E}
                \end{equation}
                must hold as well as
                \begin{equation}
                    \frac{Kc_E}{D_{I_E}} e^{-S_2(\sigma_E)} = 1.
                \end{equation}
                Again, we can see that trace preservation together with isometry necessitates that the reduced state $\sigma_E$ is pure.\\

                Additionally, as before, if we assume the state $\sigma_E = \ket{\phi_E}\bra{\phi_E}$ to be pure, we reduce the isometry condition to
                \begin{equation}
                    (\sigma_E)_I = \frac{\ibb_{I_E}}{\sum_F D_{I_F}} \qquad c_E = \frac{D_{I_E}}{\sum_F D_{I_F}}\quad ,
                \end{equation}
                which comes from the condition $|K| = \sum_E D_{I_E}$. This is, again, just the trace preservation condition. So also in the setting with nontrivial center, demanding purity makes TP and isometry equivalent. So once again, the three conditions give a 2-out-of-3 implication.

\section{Channel-State duality}\label{Ch4}
    Here, we wish to extend the usual statement of channel-state duality to the case of systems with a bipartition $(\acal_I)'=\acal_O$, but with nontrivial center in general. Therefore, we will only consider certain types of channels and states to be meaningful:
    \begin{itemize}
        \item We only consider channels $\tcal:\acal_I\rightarrow\acal_O$ between the subsystems. In terms of Hilbert space representations, this implies that the channel is block diagonal in the sectors $E$.
        \item We consider only operators ('states') which lie in the algebra generated by the union $\acal_I\cup\acal_O$. This is the set of operators generated from operations on the subsystems. 
    \end{itemize}
    These are meaningful as they preserve the intuition of channel-state duality making a statement about channels between subsystems, which correspond to states of the joint subsystems.
    The strongest kind of statement we can aim for is an isometric bijection
    \begin{equation}
        BL_k(\bcal_I,\bcal_O) \cong "\mathcal{L}_k (\hbb)" \subseteq \acal \quad ,
    \end{equation}
    identifying the k-positive linear maps between input and output algebras (the channel side) with a to-be-identified set of 'k-positive operators' given as a subset of the full algebra $\acal$ (the quotation marks indicate this lack of precise identification at this stage).\\
    The easiest motivating example is given by the Choi map, which decomposes
    \begin{equation}
        \tcal_\rho = \sum_E K c_E \tcal_{\rho_{E,E}} \quad .
    \end{equation}
    In each sector, the map $\tcal_{\rho_{E,E}}$ is a CPTP map in $\bbb(\hbb_{I,E},\hbb_{O,E})$ and therefore has a dual density matrix given by $\rho_{E,E}$. We can therefore already say that at least the usual set of k-positive operators $\mathcal{L}_k (\hbb_E)$ of each sector is possible on the right hand side:
    \begin{equation}
        \bigoplus_E \mathcal{L}_k (\hbb_E) \subseteq "\mathcal{L}_k (\hbb)".
    \end{equation}
    Also, using the same kind of Choi matrix $\tau$ as for usual channel-state-duality, we can restrict the tentative $"\mathcal{L}_k (\hbb)"$ further from above:
    Given an element $X = \sum_{E,F}X_{E,F}\in \acal$ with $X_{E,F} = \sum_{k_{E,F}} a_{k_{E,F}}\otimes b_{k_{E,F}} $, we can try to compute
    \begin{equation}
        \tau_\tcal(X) =  \sum_{E,F}\sum_{k_{E,F}} \Tr_{O_E}[\tcal(a_{k_{E,F}}) b_{k_{E,F}}^t],
    \end{equation}
    but it's clear that $\tcal(a_{k_{E,F}})$ is ill-defined unless $E=F$. Therefore, we must restrict to the sector-diagonal operators $X = \sum_E X_{E,E}$, where
    \begin{equation}
        \tau_\tcal(X) = \sum_E \tau_{\tcal,E}(X_{E,E}).
    \end{equation}
    So with both arguments, we know that the most we can expect is
    \begin{equation}
        \mathcal{L}_k (\hbb) \cong \bigoplus_E \mathcal{L}_k (\hbb_E) \quad .
    \end{equation}
    However,  the maps we consider do not mix sectors, so the statement  holds by channel-state duality in the finite dimensional, bipartite case in each sector:
    \begin{equation}
        BL_k(\bcal_I,\bcal_O) \cong \bigoplus_E \mathcal{L}_k (\hbb_E) \quad . 
    \end{equation}
    Therefore, the direct sum case of the duality reduces directly to the standard one. This is certainly not surprising, but it is worth stressing the ingredients that go into this statement. In principle, one might expect a correspondence on a much larger set of mappings on the left and operators on the right. However, under the assumptions we stated, and as our analysis of subalgebras has shown, the sets $\acal_{I|O}$ \textit{must} be chosen such that the duality becomes a per-sector statement. So in a sense, the nontrivial part lies in the selection of subsystems and the subsequent identification of the correspondence $\rho\leftrightarrow \tcal_\rho$. 
    We can also highlight the ambiguities in this correspondence: As in the single-sector case, we need to specify reference maximally entangled state for the \textit{channel $\rightarrow$ state} direction; in the current setting, this freedom is extended to a maximally entangled state \textit{per sector}. However, again, once these are chosen, a bijective isomorphism can be defined with them.
    
\section{Generalisations}\label{Ch5}
    Here we give some indications towards the generalisation of our construction and results to the infinite dimensional setting.\\
    The setting of general C*-algebras, through GNS representations and Stinespring's factorisation theorem, can essentially be reduced to the study of a cospan
    \begin{equation}
        \bbb(\hbb_I) \stackrel{V^\dagger_I(-)V_I}{\longrightarrow} \bbb(\hbb) \stackrel{V^\dagger_O(-)V_O}{\longleftarrow} \bbb(\hbb_O) \quad ,
    \end{equation}
    where all Hilbert spaces are separable and the maps $V_{I|O}:\hbb\rightarrow \hbb_{I|O}$ are bounded. 
    We can generalize our results by taking a more general cospan of bounded operator algebras between Hilbert spaces. 
    More precisely, let us list the key ingredients in our construction and how each of them can be generalised.
    \begin{enumerate}
        \item Existence of states $\rho$. This can be generalised to density matrices in the folium of a state $\omega\in \mathcal{S}(\acal)$ of some C*-algebra.
        \item Existence of the identities $\ibb$ for use in defining extension maps. They may be generalised to approximate identities of C*-algebras and associated nets of approximate extension maps.
        \item Existence of partial traces. Relative to an extension map, one may either take the inverse or the adjoint generalisation. Working with the inverse is maybe simpler and requires less structure. For the adjoint variant, we can either take the Banach adjoint or, in the presence of a scalar product, use the associated Hilbert adjoint (if the partial trace is bounded or at least densely defined). Either choice, though, will generically force us to move away from the full algebra and at least into the space of compact operators, as even the identity $\ibb$ does not have finite trace on infinite dimensional spaces.
    \end{enumerate}
    Recall first some relevant notation and the subsets of bounded operators on an infinite dimensional Hilbert space $\hbb$:
    \begin{equation}
        \text{Fin}(\hbb)  \subset L^1(\hbb) \subset L^2(\hbb) \subset K(\hbb) \subset \bbb(\hbb) \quad .
    \end{equation}
    These are the finite rank, trace class, Hilbert-Schmidt, compact and bounded operators, respectively. Hilbert-Schmidt rank ones obey the special property $L^2(\hbb) \cong \hbb^\ast\otimes \hbb$. In this notation, we will find a generalised mapping between
    \begin{equation}
        \tcal: \bbb(\hbb_I) \rightarrow L^1(\hbb_O) \quad ,
    \end{equation}
    which may be restricted to a mapping on Hilbert-Schmidt operators.\\
    
    We propose here one particular generalisation for the bipartite case, focused on trace class operators. 
    Given some bounded injective *-homomorphism between the (non-unital!) C*-algebra of compact operators $K(\hbb_O)\stackrel{i_O}{\rightarrow} K(\hbb)$, we can form its Banach adjoint\footnote{The usual issues of domains apply, but as long as the extensions are bounded, we may neglect them.}
    \begin{equation}
        \Tilde{i}_O: K(\hbb)^\ast\rightarrow K(\hbb_O)^\ast, \Tilde{i}_O(f)(x) = f(i_O(x)) \quad .
    \end{equation}
    Then, we use the fact that the map
    \begin{equation}
        \alpha_K: L^1(\hbb)\rightarrow K(\hbb)^\ast, \alpha_K(x)(y) = \Tr[xy] \quad ,
    \end{equation}
    turning trace class operators into associated functionals, is an isometric bijection (see Proposition 3.4 of\cite{muger_trace_2022}). We can then define a partial trace as
    \begin{equation}
        P\Tr_O = \alpha^{-1}_K \circ \Tilde{i}_O \circ \alpha_K : L^1(\hbb)\rightarrow L^1(\hbb_O) \quad , 
    \end{equation}
    which satisfies
    \begin{equation}
        \begin{gathered}
            \Tr[P\Tr_O(x) y] = \Tr[x i_O(y)] \\ 
            \forall \, y \in K(\hbb_O), x\in L^1(\hbb) \quad .
        \end{gathered}
    \end{equation}
    Note that these extension and partial trace operations do not have compatible (co)domains, as the partial trace is only defined on a subset of the target. Therefore, we will also require that our extension maps $i_{I|O}$ satisfy
    \begin{equation}
        i_{I|O}(L^1(\hbb_{I|O})) \subseteq L^1(\hbb) \quad .
    \end{equation}
    With this in hand, we can proceed as in the finite-dimensional case. Given an extension map $\bbb(\hbb_I)\stackrel{i_I}{\rightarrow} \bbb(\hbb)$, we can twist it by some $\Sigma: \bbb(\hbb)\rightarrow L^1(\hbb)$ and get
    \begin{equation}
        \tcal_\Sigma: \bbb(\hbb_I)\stackrel{i_I}{\rightarrow} \bbb(\hbb) \stackrel{\Sigma}{\rightarrow} L^1(\hbb) \stackrel{P\Tr_O}{\rightarrow} L^1(\hbb_O) \quad ,
    \end{equation}
    which is the analogue of our transport superoperator from the finite-dimensional case. 
    In particular, we may choose
    \begin{equation}
        \Sigma(x) = \rho^{(t_I)} x, \rho \in L^1(\hbb) \quad ,
    \end{equation}
    which gives the Jamiolkowski or Choi map, depending on whether we use the transpose or not. We note that this construction must be a special case of the general fact\cite{majewski_comment_2013} that 
    \begin{equation}\label{IsomIsom}
        BL(\bbb(\hbb_I),\bbb(\hbb_O)) \stackrel{\text{isometric}}{\cong} (\bbb(\hbb_I)\otimes_\pi L^1(\hbb_O))^\ast \quad , 
    \end{equation}
    in which mappings between operator algebras are equivalent to linear functionals on a projective tensor product algebra. We have here a case 
    \begin{equation}
       L^1(\hbb_I\otimes\hbb_O) \rightarrow BL(\bbb(\hbb_I),L^1(\hbb_O)) \quad .  
    \end{equation}
    So, in order to connect to the previous work, one should find a map $L^1(\hbb) \rightarrow (\bbb(\hbb_I)\otimes_\pi L^1(\hbb_O))^\ast$. This must be
    \begin{equation}
        \rho \mapsto \tau_\rho, \, \tau_\rho(x\otimes y) = \Tr[(i_I(x) i_O(y)^{t_O})\rho^{(t_I)}] \quad ,
    \end{equation}
    simply using the expression of the isometric isomorphism \ref{IsomIsom}. This functional is defined on $\bbb(\hbb_I)\otimes_\pi K(\hbb_O)$, which what we needed. Our example fits directly into a small class of maps in the general scheme which can be represented by a density matrix $\rho$. Using the general statement of isometry, we even have a way to estimate the norm of $\tcal$:
    \begin{equation}
        ||\tcal|| = \sup_{v\in \bbb(\hbb_I)\otimes_\pi L^1(\hbb_O)} \, \frac{|\tau_\rho(v)|}{\pi(v)} \quad ,
    \end{equation}
    with $\pi$ the projective norm on $\bbb(\hbb_I)\otimes_\pi L^1(\hbb_O)$, under which it is a Banach space. For it to be an isometry, of course, we need that $||\tcal|| = 1$. In this infinite dimensional setting, this is the most we can require as there is no good notion of, say, the partial trace in the Hilbert-Schmidt setting directly.\\
    Now, granted the validity of the general scheme, we still need concrete injection maps. 
    The central issue to be tackled in the construction is not $i_I$, which we can simply take to be $i_I(x) = x\otimes_\pi \ibb_O$, but rather $i_O$, which can not work in the same way as in the finite-dimensional case, since $\ibb_I$ is not compact. Instead, we must content ourselves with an approximate identity
    \begin{equation}
        \{\ibb^{(\lambda)}_I \in K(\hbb_I) \,| \, \ibb^{(\lambda)}_I \text{ s.a., }   \lambda \in \Lambda\}
    \end{equation}
    and thus construct an approximate family of injections $i^{(\lambda)}_O(y) = \ibb^{(\lambda)}_I\otimes y $.
    This, then, results in a family of superoperators whose formal limit
    \begin{equation}
        \tcal = \lim_{\lambda \in \Lambda} \tcal^{(\lambda)} = \lim_{\lambda \in \Lambda} (P\Tr^{(\lambda)}_O \circ \Sigma \circ i_I)
    \end{equation}
    gives us the mapping we seek. Whether this limit exists as an operator in $BL(\bbb(\hbb_I),L^1(\hbb_O))$ is of course nontrivial, but it is a requirement for $\tcal$ to be an isometry.\\
    Now if we also restrict the mapping to Hilbert-Schmidt operators, $L^2(\hbb_I)$, then we can in fact still speak of the same kind of isometry under Hilbert-Schmidt:
    \begin{equation}
        \avg{\tcal(X),\tcal(Y)}_{L^2(\hbb_O)} \stackrel{?}{=} \avg{X,Y}_{L^2(\hbb_I)} \quad .
    \end{equation}
    Note, though, that the Banach adjoint exists here, while the Hilbert adjoint is not defined (as $L^1(\hbb_O)$ is not a Hilbert space). We therefore will need to make a choice on what kind of isometry we are looking for - or rather, which norm we wish to preserve.\\
    
    \section{Conclusions}
    We have studied information transport channels for quantum systems defined by operator algebras with centers, corresponding to Hilbert spaces with a structure of a direct sum of factorized Hilbert spaces (each identifying a natural notion of subsystems), generalizing the usual situation of factorized Hilbert spaces. Operator algebras and Hilbert spaces of this type appear in a very broad and diverse range of physical contexts, from condensed matter and quantum many-body systems, to (lattice) gauge theories, to quantum gravity and holography. In particular, we have proposed a generalization of the usual channel-state duality adapted to this direct sum context. We have also sketched how our construction, given in the finite-dimensional case, could be extended to the infinite-dimensional one. These transport channels offer a powerful way to characterize and study quantum correlations (entanglement) beyond the simple definition relying on non-separability of quantum states. This conceptual and analytic power, together with the vast range of physical systems with the algebraic characterization we considered, imply the possibility of many future developments from the application of our construction.

\section*{Acknowledgements}
SL would like to thank D. Aliverti and L.Csillag for valuable feedback on the manuscript and useful discussions. SL and DO acknowledge funding from the Munich Center for Quantum Science and Technology. DO also acknowledges funding from the Deutsche Forschungsgemeinschaft (DFG), and the ATRAE programme of the Spanish Government, through the grant PR28/23 ATR2023-145735. ECs participation in this project was made possible by a DeBenectis Postdoctoral Fellowship and through the support of the ID\# 62312 grant from the John Templeton Foundation, as part of the \href{https://www.templeton.org/grant/the-quantum-information-structure-of-spacetime-qiss-second-phase}{‘The Quantum Information Structure of Spacetime’ Project (QISS)}.

\bibliographystyle{unsrt}
\bibliography{P1S1}

\end{document}